\pdfoutput=1
\documentclass{article}
\usepackage{spconf,amsmath,graphicx,hyperref}
\usepackage{xcolor}
\usepackage{subcaption}

\title{Enhancing Conversational TTS with Cascaded Prompting and ICL-Based Online Reinforcement Learning}

\makeatletter
\def\@name{{\em Zhicheng Ouyang, Seong-Gyun Leem, Bach Viet Do, Haibin Wu,}\\
{\em Ariya Rastrow, Yuzong Liu, Florian Metze}\\}
\makeatother
\address{Meta AI, USA}

\begin{document}
\maketitle

\begin{abstract}
Conversational AI has made significant progress, yet generating expressive and controllable text-to-speech (TTS) remains challenging. Specifically, controlling fine-grained voice styles and emotions is notoriously difficult and typically requires massive amounts of heavily annotated training data. To overcome this data bottleneck, we present a scalable, data-efficient cascaded framework that pairs textual style tokens with human-curated, high-quality audio prompts. This approach enables single-shot adaptation to fine-grained speaking styles and character voices. In the context of TTS, this audio prompting acts as In-Context Learning (ICL), guiding the model's prosody and timbre without requiring massive parameter updates or large-scale retraining. To further enhance generation quality and mitigate hallucinations, we introduce a novel ICL-based online reinforcement learning (RL) strategy. This strategy directly optimizes the autoregressive prosody model using subjective aesthetic rewards while being constrained by Connectionist Temporal Classification (CTC) alignment to preserve intelligibility. Comprehensive human perception evaluations demonstrate significant improvements in both the naturalness and expressivity of the synthesized speech, establishing the efficacy of our ICL-based online RL approach.
\end{abstract}


\section{Introduction}
\label{sec:intro}
Conversational AI has made remarkable progress, yet generating expressive and controllable text-to-speech (TTS) remains a significant challenge. Conversational audio large language models (LLMs), for instance, often struggle to control voice expressivity due to the limited availability of expressive conversational audio and the absence of reliable reward models for alignment. To overcome these limitations, we explore a text-native cascaded ASR–LLM–TTS paradigm, which leverages the inherent controllability of LLMs to unlock the expressive potential of conversational TTS. A major hurdle in expressive TTS is the precise control of fine-grained voice styles and emotions, which conventionally demands massive datasets of heavily annotated emotional speech. To bypass this data-intensive requirement, we propose a scalable, data-efficient framework that enables single-shot adaptation to fine-grained speaking styles and character voices. Our key insight is that providing a short, high-quality audio clip as a reference prompt to the TTS model constitutes a form of In-Context Learning (ICL): the model adapts its output style at inference time without any weight updates, guided solely by the audio context paired with a textual style token. We exploit this property at two levels of a cascaded architecture—an autoregressive (AR) prosody model and a diffusion-based acoustic model—to achieve fine-grained, consistent expressivity across multi-turn conversations. Furthermore, to enhance generation quality, we introduce a novel ICL-based online RL strategy. Traditional posterior sampling methods select the best sample from multiple candidates at inference time, which is computationally expensive. Instead, our approach directly optimizes the AR prosody model during training using a subjective aesthetic quality reward (AES-CE), while employing CTC loss as a regularizer to prevent reward hacking and text hallucinations. Extensive human evaluations demonstrate substantial improvements in both naturalness and expressivity, underscoring the effectiveness of combining cascaded prompting with ICL-based online RL.

\section{Related Work}
\label{sec:related}

In recent years, the field of expressive and controllable TTS has seen remarkable progress, emerging as a vibrant and significant area of research.
One category in emotional TTS was characterized by a coarse-grained approach~\cite{guo2023emodiff,zhou2022speech}, relying on discrete category labels to control expression. These systems primarily focused on synthesizing a set of mainstream emotions, such as happiness, sadness, and anger.
Subsequently, the field evolved towards fine-grained emotional control by adopting dimensional models~\cite{wu2024laugh,cho2025emosphere++}. This paradigm leverages continuous representations—typically Valence (the pleasure-displeasure axis) and Arousal (the intensity-activation axis), sometimes extended to include Dominance (control)—as signals to precisely modulate the synthesized speech.
Recent works have pushed the boundaries further by using fine-grained natural language descriptions as emotional guidance~\cite{yang2025emovoice}. This cutting-edge approach allows for an unprecedented level of specificity, enabling the synthesis of complex and nuanced emotional states that are difficult to capture with predefined labels or dimensional values alone.

Conversational TTS has become a prominent research area, focused on generating speech that sounds natural and dynamically interactional.
A prevalent architectural approach involves a two-stage pipeline: first, an autoregressive (AR) model predicts a sequence of semantic tokens from the input text, capturing the conversational flow and dependencies. Subsequently, a non-autoregressive (NAR) model rapidly generates the final acoustic features from these tokens. This hybrid design is exemplified by representative models like MoonCast~\cite{ju2025mooncast} and CoVoMix~\cite{zhang2024covomix}.
In contrast, an alternative strategy employs a single, end-to-end generative model to synthesize the entire dialogue directly from the text, typically leveraging powerful frameworks like flow-matching or diffusion. Works such as ZipVoice-Dialog~\cite{zhu2025zipvoice} and CovoMix2~\cite{zhang2025covomix2} adopt this method.
Another approach, demonstrated by NotebookLM~\cite{deepmindNotebookLM} and Sesame-CSM~\cite{sesameCrossingUncanny}, utilizes a hierarchical transformer to autoregressively produce a continuous stream of audio tokens, which are then decoded into the final dialogue waveform.

To further improve the expressivity and naturalness of TTS, researchers have investigated reinforcement learning (RL) approaches, where the model (or policy) is updated online in response to reward signals~\cite{liu2021reinforcement, chen2024rltts}. Online RL methods leverage feedback from sources such as emotion detectors, UTMOS predictors, and diffusion-based losses to enhance emotion discriminability and perceptual speech quality. Complementing these efforts, offline RL strategies have also been explored. For example, Gao et al.~\cite{gao2024emo-dpo} proposed an emotional DPO training approach to better align TTS models with human emotion preferences.

\section{Proposed Approach}
\label{sec:typestyle}
To fully harness the expressive potential of conversational TTS, we propose a cascaded framework that enables contextual and scalable control of voice expressivity through the integration of \emph{textual style tokens} and \emph{audio prompts}. The overall paradigm is illustrated in Figure 1. In this framework, the primary expressivity control signal passed from the LLM to the TTS module is the textual style token: it is first generated by the LLM based on conversational context, then extracted and forwarded to the TTS system. Notably, this design is fully compatible with real-time AI systems.
\begin{figure}[htb]
  \centering
  \centerline{\includegraphics[width=8.5cm]{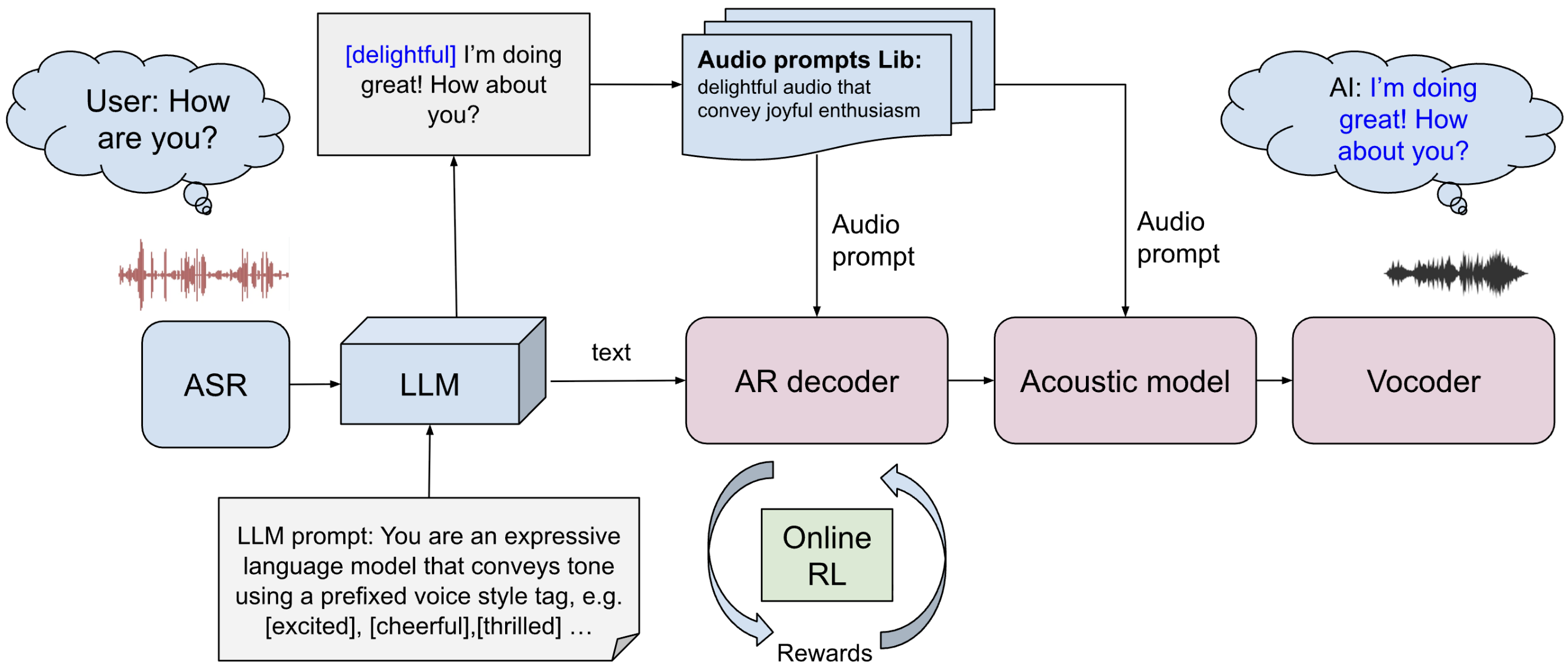}}
\caption{Cascaded conversational framework combining expressive LLM prompts with audio style prompts, enhanced through online reinforcement learning to improve speech generation quality.}
\label{fig:framework}
\end{figure}

\subsection{Cascaded Prompting}
\label{sec:cascaded_prompting}
Textual style tokens are contextually generated by an LLM based on expressive text prompts and the surrounding conversational context. The role of audio prompts is to align the synthesized speech with the LLM's predicted style token. To achieve this, human listeners select a representative audio prompt for each style token by considering the paralinguistic dimensions of prosody, valence, arousal, and dominance. Although prompt selection involves human-in-the-loop listening, the approach remains highly scalable and data-efficient: rather than requiring massive datasets of emotional speech for training, each fine-grained style token requires only a single high-quality audio prompt to effectively convey its expressivity via In-Context Learning.
\subsubsection{Autoregressive Prosody Prompting}
\label{ssec:lm_prompt}
Our autoregressive (AR) prosody model is pretrained on a large corpus of human speech and generates discrete tokens that capture the prosody of any target voice style. To ensure high-quality and stylistically accurate synthesis, we curate a pool of candidate audio prompts for each style and generate at least ten speech samples per candidate. Each sample is evaluated using the Aesthetic Quality Score focusing on Content Enjoyment (AES-CE), a perceptual metric that correlates with human preferences for acoustic quality. We establish a lower-bound score threshold for each candidate using Monte Carlo estimation and select the candidate with the highest lower-bound score. A final human-in-the-loop listening validation step filters out candidates that produce undesirable artifacts, such as hallucinations or unnatural breathing.

\subsubsection{Diffusion-based Acoustic Prompting}
\label{ssec:am_prompt}
Rather than reusing the same audio prompt from the AR prosody model for every style, we propose reducing the granularity of styles at the acoustic modeling stage. This decision is motivated by observed \emph{speaker drift} in multi-turn conversations: even prompts recorded from the same speaker can vary in volume, spatial characteristics, and timbre across recording sessions. Figure 2 illustrates how speaker similarity varies with style granularity. We find that grouping fine-grained styles into coarser categories for acoustic prompting substantially reduces speaker drift. A further observation is that the audio prompt speaker used in the AR prosody model need not match the speaker used in the acoustic model prompt. Voice timbre is predominantly determined by the acoustic model, while the AR model primarily controls prosody. This finding elegantly \emph{decouples prosody and timbre control}, allowing independent optimization of each dimension.
\begin{figure}[htb]
  \centering
  \centerline{\includegraphics[width=8.5cm]{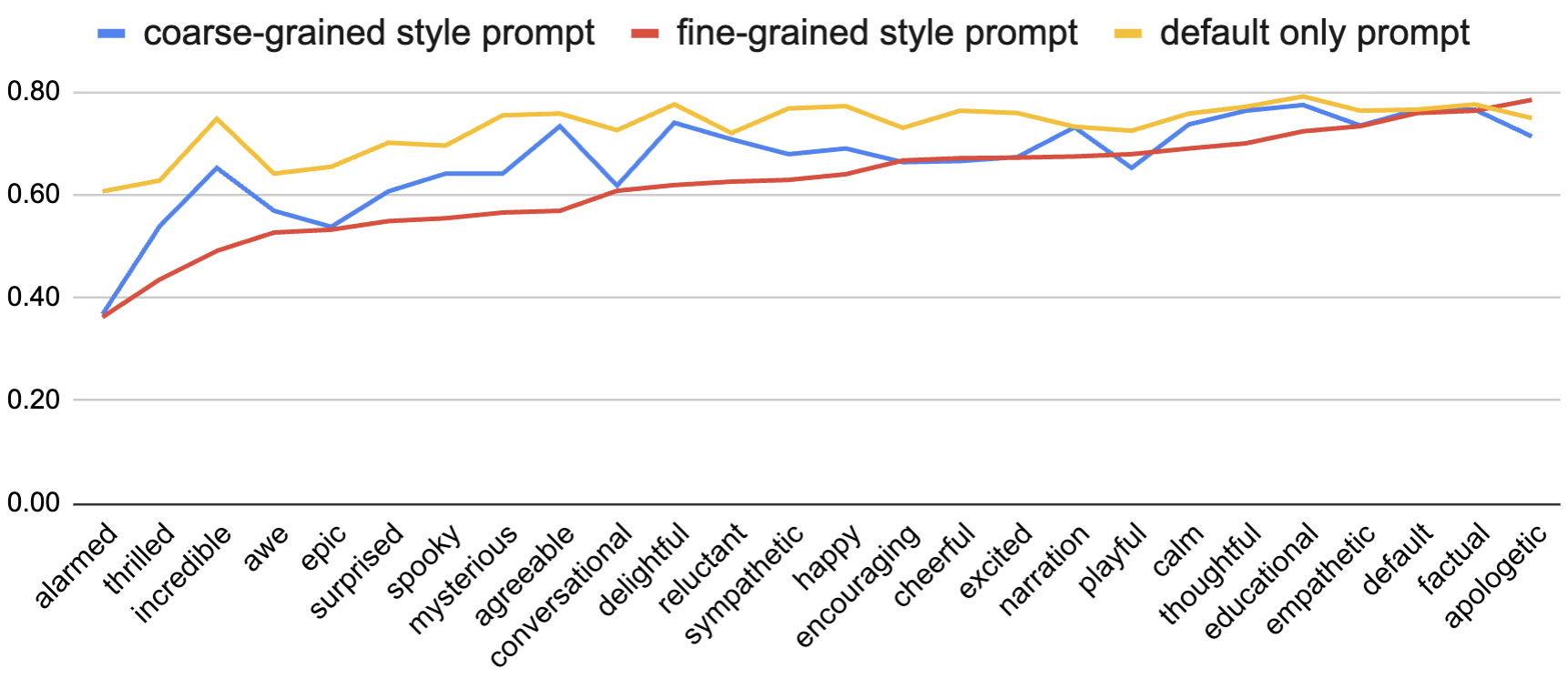}}
\caption{Average speaker similarity score variations with respect to different style granularities in acoustic modeling.}
\label{fig:speaker_sim}
\end{figure}

\subsection{Generation Quality Evaluation Protocol}
\label{sec:style_quality}
To evaluate the expressivity of synthesized voices, we designed a rating protocol based on four paralinguistic dimensions: \textbf{C}larity, \textbf{V}alence, \textbf{A}rousal, and \textbf{D}ominance (CVAD). Raters assessed the appropriateness of each dimension and overall expressivity on a five-point Likert scale (1 = very inappropriate, 5 = very appropriate). Comparative Mean Opinion Score (CMOS) values were computed from the overall expressivity ratings by comparing different AI voices on the same conversational context. For multi-turn scenarios, we additionally measure \emph{speaker consistency} using the ECAPA-TDNN speaker verification model. Each generated sample's speaker embedding is compared to a reference target embedding (the average of a curated pool of target speaker embeddings). A cosine similarity threshold of 0.7 is used to flag speaker drift, ensuring the synthesized voice remains recognizable throughout the conversation.

\subsection{ICL-Based Online Reinforcement Learning}
\label{sec:online_rl}

We envision online RL training as a more efficient alternative to
posterior sampling-based re-ranking. Rather than using rewards to select
the best sample from multiple candidates at inference time, we apply the
reward \emph{directly during training} to guide the AR prosody model
toward generating the most aligned output. Crucially, the RL policy is
conditioned on the same audio prompts used during ICL inference, making
this an \emph{ICL-based} online RL approach: the model learns to
generate better speech in context, not just in isolation.

We use AES-CE as the primary reward, as it correlates well with human
preferences for acoustic quality. However, optimizing solely for AES-CE
leads to reward hacking, manifesting as severe text hallucinations. To
address this, we incorporate CTC loss~\cite{graves2006ctc} into the
reward function, which aligns the generated audio token sequences with
the ground-truth transcript. The combined reward function is:
\begin{equation}
  R(\tau) = \alpha_\text{AES} \cdot \text{AES}\!\left(\mathcal{F}(\tau)\right)
            - \alpha_{\text{CTC}} \cdot \mathcal{L}_{\text{CTC}}(\tau, w_0)
  \label{eq:reward}
\end{equation}
where $\tau$ denotes the discrete audio tokens generated by the AR
prosody model for transcript $w_0$, and $\mathcal{F}(\cdot)$ denotes the
frozen acoustic model and vocoder pipeline that converts audio tokens to
a waveform. The hyperparameters $\alpha_\text{AES}$ and
$\alpha_{\text{CTC}}$ balance the two reward components. The RL
objective is to maximize:
\begin{equation}
  J(\theta) = \mathbf{E}_{\tau \sim \pi_\theta}[R(\tau)]
              - \beta \cdot \mathrm{KL}(\pi_{\theta} \| \pi^{0})
  \label{eq:rl_obj}
\end{equation}
where $\pi_\theta$ is the online (trainable) policy of the AR model,
$\pi^0$ is the reference policy (the Supervised Fine-Tuned (SFT)
baseline), and $\beta$ controls the KL divergence penalty to prevent the
policy from deviating too far from the reference.

\section{Experiments}
\label{sec:experiments}
\textbf{Model baseline}: Our baseline models consist of Meta's LLaMA 3 70B model~\cite{meta_llama3_70b} paired with an in-house TTS system that follows a modeling architecture similar to Tortoise-TTS~\cite{betker2023better}, where the vocoder is implemented using BigVGAN.

\noindent\textbf{Dataset}: We used in-house voice actor data for prompt candidates, with coarse-grained voice styles provided by the vendor. Based on these coarse-grained style audios, we conducted listening judgments and re-mapped a few coarse-grained candidates into each fine-grained voice style. The final prompts were chosen through the manual process described in Section~\ref{sec:cascaded_prompting}.

\subsection{Prompt Selection with Humans in the Loop}
\label{sec:exp_prompt}
To enhance the expressivity and naturalness of conversational TTS, we investigated the advantage of cascaded prompting, referred to as the in-context learning (ICL) model pipeline, over the default setting without ICL (i.e., missing prompt in the decoder and using a default prompt in the acoustic model). For the ICL model pipeline, we used the TTS model with fine-grained style prompts described in Section~\ref{sec:cascaded_prompting}. For evaluation, we generated 400 voice samples for each of 10 speakers (7 with US English accents and 3 with British English accents) under both Zero-shot baseline and ICL conditions. Fifty crowd raters assessed the naturalness of these samples using CMOS, which quantifies the net win ratio between models. To provide a more targeted assessment of expressivity, we employed the CVAD framework outlined in Section~\ref{sec:human_judge}. For this evaluation, we generated 50 samples from each model and collected ratings from five expert raters per sample. In addition to comparing the ICL and Zero-shot baseline models, we also conducted a study comparing our models with GPT-4o's external API~\cite{Hurst_2024}.

\begin{table}[h!]
\centering
\caption{CMOS of ICL-setting TTS over Baselines Across Tasks}
\label{tab:icl_cmos}
\noindent \resizebox{\columnwidth}{!}{%
\begin{tabular}{|c|c|c|c|}
\hline
\textbf{Task} & \textbf{Model} & \textbf{Baseline} & \textbf{Net Win Rate (\%)} \\
\hline
Naturalness & ICL & Zero-shot baseline & +7.5 \\
CVAD & ICL & Zero-shot baseline & +79.6 \\
CVAD & ICL & GPT4o & +5.6 \\
\hline
\end{tabular}
}
\end{table}

Table~\ref{tab:icl_cmos} presents our subjective evaluation results, highlighting the clear advantage of the cascaded prompting approach. The ICL pipeline achieved a +7.5\% net win rate in naturalness CMOS over the Zero-shot baseline baseline, indicating a substantial improvement in perceived naturalness. To assess expressivity, we conducted a human evaluation using the CVAD framework (Section~\ref{sec:human_judge}) on an emotional wellness dataset of 50 samples. Results show that the ICL model outperformed the Zero-shot baseline baseline by +79.6\% in CVAD CMOS and even surpassed GPT-4o by +5.6\%. While our evaluation is not based on a standardized benchmark, the substantial expressivity gains remain valid and are largely attributed to the effective human-in-the-loop alignment employed during audio prompt selection. Overall, these findings demonstrate that prompt tuning with in-context learning not only improves the naturalness of TTS outputs but also yields more expressive and contextually appropriate speech, as confirmed by comprehensive human evaluations.

\subsection{ICL-Based Online RL Training}
\label{sec:exp_rl}
For RL training, we constructed an audio-conditioned transcript dataset consisting of pairs of audio prompts and target transcripts, where the content of the target transcript is independent of the audio content. During training, the acoustic model and vocoder were frozen, and only the AR prosody model was updated. For training efficiency, we reused the same audio prompts as in the inference setting. To approximate the term $ \mathbf{E}_{\tau \sim \pi_\theta(\cdot)}[R(\tau)]$, we applied a Monte Carlo estimate with six samples per transcript. The key training hyperparameters were set as follows: $\alpha_{\text{AES}} = 1.0$, $\alpha_{\text{CTC}} = 3.0$, and $\beta = 50.0$. The Adam optimizer was used with a learning rate of $5.0 \times 10^{-7}$.

\begin{figure}[h]
    \centering
    \begin{subfigure}[t]{0.45\linewidth}
        \centering
        \includegraphics[scale=0.4]{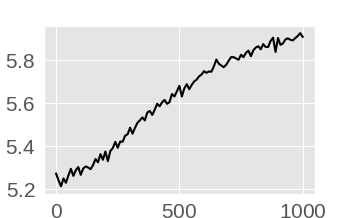}
        \caption{AES-CE training score}
        \label{fig:rl_aes}
    \end{subfigure}
    ~
    \begin{subfigure}[t]{0.45\linewidth}
        \centering
        \includegraphics[scale=0.4]{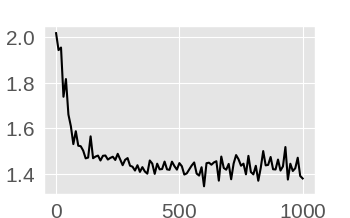}
        \caption{CTC training loss}
        \label{fig:rl_ctcloss}
    \end{subfigure}
    ~
    \begin{subfigure}[t]{0.45\linewidth}
        \centering
        \includegraphics[scale=0.4]{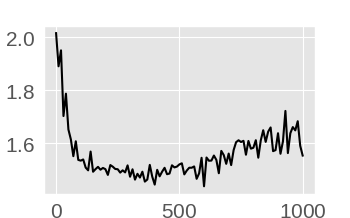}
        \caption{CTC training loss without enabling it for backpropagation.}
        \label{fig:rl_ctcloss2}
    \end{subfigure}
    \caption{Loss and Scores vs RL Training Iterations}
    \label{fig:rl_training}
\end{figure}

Figure~\ref{fig:rl_training} provides an overview of the training process. Figure~\ref{fig:rl_aes} shows that the AES-CE score steadily increased, while hallucinations were suppressed, as indicated by the controlled level of CTC loss in Figure~\ref{fig:rl_ctcloss}. For comparison, when CTC loss was not enabled during training, we recorded the resulting CTC loss in Figure~\ref{fig:rl_ctcloss2}. Regarding human evaluation results, the RL-AES-CTC model after SFT outperforms the baseline model trained with SFT only, showing an improvement of approximately $+7\%$ in terms of CMOS.

\begin{table}[h]
    \centering
    \caption{Human CMOS evaluation results between RL enhanced version and baseline SFT only version. \\}
    \label{tab:my_example_table}
\noindent \resizebox{\columnwidth}{!}{%
    \begin{tabular}{|c|c|c|}
      \hline
      \textbf{Ablation} & \textbf{Average winrate} & \textbf{ 95\% Confidence Interval} \\
      \hline
      RL-AES-CTC vs SFT Only  & $+7.1\%$ & $(3.97\%,10.23\%) $  \\
      \hline
    \end{tabular}
}
\end{table}
\section{Results}
\label{sec:results}
\subsection{Human judge}
\label{sec:human_judge}
We have observed significant improvements in perceptual quality through our cascaded prompting strategy and AES-CE–guided online RL. Although our TTS system and prompt data are entirely in-house, we believe that a similar TTS model architecture, combined with a pretrained decoder and commonly available expressive human speech data, could achieve comparable improvements.
\subsection{Audio demo}
\label{sec:demo}
We present single-turn and multi-turn conversation examples featuring a variety of voice styles, illustrating results before and after applying our cascaded prompting approach. The examples are available on our \href{https://incontts.github.io/InconTTSDemo/}{demo webpage} (\url{https://incontts.github.io/InconTTSDemo/}).

\section{Conclusion}
\label{sec:conclusion}
We presented a data-efficient, cascaded conversational TTS framework that supports single-shot adaptation to fine-grained speaking styles and character voices. By pairing textual style tokens with human-curated audio prompts, we overcome the traditional bottleneck of requiring massive emotional speech datasets, leveraging In-Context Learning (ICL) to achieve precise expressivity control. To further enhance generation quality and address alignment issues, we introduced a novel ICL-based online reinforcement learning strategy. By directly optimizing the autoregressive prosody model with aesthetic rewards while regularizing with CTC loss, our approach effectively mitigates hallucinations and reward hacking. Extensive human evaluations demonstrate that our framework achieves superior naturalness and expressivity compared to zero-shot baselines and SFT-only models, highlighting its potential for scalable, expressive conversational AI.

\bibliographystyle{IEEEbib}
\bibliography{refs}

\end{document}